\documentstyle[12pt,epsf]{article}
\begin{document}

\title{On the notion of potential in quantum gravity}

\author{~Kirill~A.~Kazakov\thanks{E-mail: $kirill@theor.phys.msu.su$}}

\maketitle

\begin{center}
{\em Moscow State University, Physics Faculty,\\
Department of Theoretical Physics.\\
$117234$, Moscow, Russian Federation}
\end{center}

\begin{abstract}
The problem of consistent definition of the quantum corrected gravitational
field is considered in the framework of the $S$-matrix method.
Gauge dependence of the one-particle-reducible part of the
two-scalar-particle scattering amplitude, with the help of which the
potential is usually defined, is investigated at the one-loop approximation.
The $1/r^2$-terms in the potential, which are of zero
order in the Planck constant $\hbar,$ are shown to be independent
of the gauge parameter weighting the gauge condition in the action.
However, the $1/r^3$-terms, proportional to $\hbar,$ describing the first
proper quantum correction, are proved to be gauge-dependent. With the help
of the Slavnov identities, their dependence on the weighting parameter
is calculated explicitly. The reason the gauge dependence originates from
is briefly discussed.
\end{abstract}
PACS number(s): 04.60.Ds, 11.15.Kc, 11.10.Lm


\noindent

\section{Introduction}

Quantization of the General Theory of Relativity is conventionally
performed along the formal lines of quantization of the ordinary Yang-Mills
theories. Apart from complications introduced by the gauge invariance, 
both are carried out on the basis of Bohr's correspondence principle 
that gives certain prescriptions as to
construction of the operators for physical field quantities.
It implies, in particular, that the non-commutativity of these operators
becomes negligible when the occupation numbers of physical states get large,
and so the quantum equations of motion of free fields become effectively
classical. Switching on the interaction results in both the classical
nonlinear and quantum radiative corrections to these equations.
The property of being classical, however, should be retained by the
largely occupied states even in the presence of interaction, at least
in the case of small coupling constants (or small time intervals the
states are observed in). The radiative corrections to these states are
thus supposed to be measurable in the classical sense, since it is the
filling of states, rather than the relative value of the corrections,
that determines the system property of being classical.

As is well known, the above immediate interpretation of the effective
fields runs into the problem of their gauge dependence. One is prompted
therefore to seek an indirect interpretation based on the use of explicitly
gauge-independent means.

In many cases, a gauge-independent definition of the potential can be
given with the help of the $S$-matrix which gauge-independence
is insured by the well-known equivalence theorem \cite{dewitt1,ktut}.
In the case of spinor electrodynamics, for instance, the potential
can be defined with the help of the two-particle scattering amplitude
Fourier transformed with respect to the momentum transfer between the
particles. Incidentally, with the help of the potential so defined one
usually formulates the physical renormalization conditions which are
nothing but the classical definitions of the charges and masses of the
particles.

There is, however, an obstacle in direct application of the equivalence
theorem to the potential. The point is that the latter cannot be
defined directly through the two-particle scattering amplitude, since the
set of Feynman graphs describing given scattering process contains
irreducible as well as reducible with respect to the gauge field diagrams.
Only after the reducible part is separated out of the whole set of diagrams
can the notion of the potential be introduced by a straightforward
generalization of the usual definition used in electrodynamics.
This is exactly the way followed in Ref.~\cite{donoghue}
in investigation of the post-Newtonian classical and quantum corrections
to the gravitational potential.

The purpose of this paper is to investigate consistency of the
above-mentioned separation in the case of quantum gravity.
As will be explained in Sec.~\ref{definition},
actually there is no intrinsic reason underlying the division of diagrams
according to the property of reducibility in this case, threatening thereby
validity of the equivalence theorem as applied to the reducible subset of
diagrams. That the potential defined with the help of this subset does
depend on the gauge, loosing thereby any significance as a means for
description of particle interactions, is shown in Sec.~\ref{loop} by an
explicit calculation. Sec.~\ref{tools} contains an account of the method
used in evaluation of the gauge-dependence of the one-loop logarithmic
radiative corrections. The results of the work are discussed in
Sec.~\ref{conclude}. Some formulae needed in calculation of the Feynman
integrals are obtained in the Appendix.

The highly condensed notations of DeWitt \cite{dewitt1} are employed
throughout this paper. Also left derivatives with respect to anticommuting
variables are used. The dimensional regularization of all divergent
quantities is supposed.

\section{Definition of the potential in quantum gravity}\label{definition}

It was mentioned in the Introduction that the notion of potential does
make sense only if one is justified to disregard the set of Feynman
graphs irreducible with respect to the gauge field. Before we proceed
to actual calculations, let us consider this point in more detail.

Note, first of all, that the potential must be defined in terms
characterizing motion of interacting particles, simply because only in
this case the definition would be relevant to an experiment.
For this purpose, the scattering matrix approach can
be used, in which case the potential is conventionally defined as the
Fourier transform (with respect to the momentum transfer from one particle
to the other) of the suitably normalized\footnote{The normalization is
fixed by the requirement that the potential takes the Newtonian form at
the tree level.} two-particle scattering amplitude.
By itself this definition is not of great value unless one is able to
separate the whole scattering process as follows: interaction of the
first particle with the gauge field $\to$ propagation of the gauge field
$\to$ interaction of the gauge field with the second particle.
Only if such a separation is possible can one introduce a self-contained
notion of the potential. In terms of the Feynman diagrams, one would say
in this case that the diagrams describing the scattering process are
one-particle-reducible with respect to the gauge field.

\begin{figure}
\epsfxsize=15,5cm\epsfbox{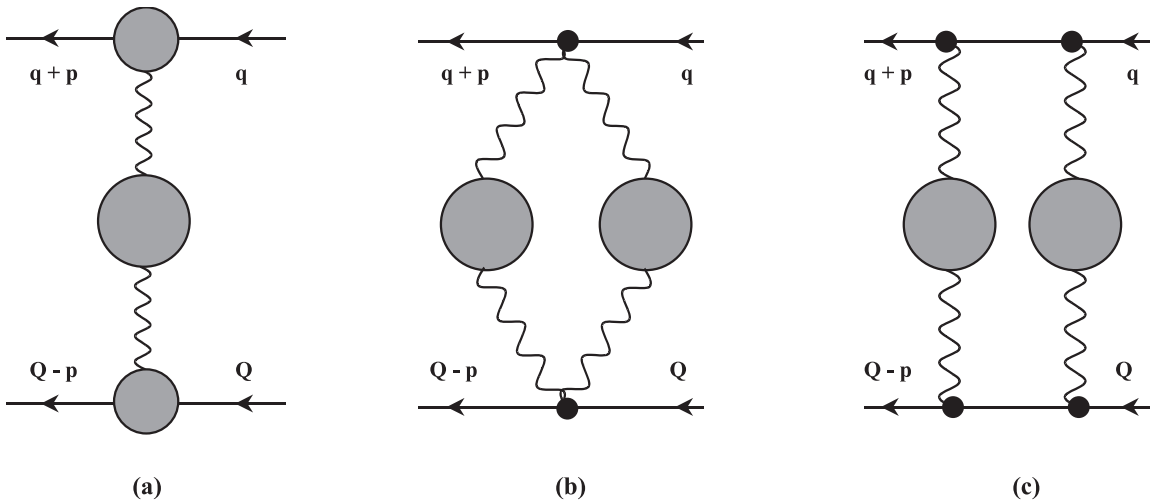}
\vspace{1cm}
\caption{Feynman graphs representing general structure of various
contributions to the two-particle scattering amplitude. (a) The
one-particle-reducible part. (b) Contributions occurring
when the gauge field -- matter interaction is nonlinear in the gauge field.
(c) The irreducible contribution to the gravitational scattering amplitude,
remaining finite in the limit $m \to \infty.$ Wavy lines represent
gravitons, solid lines matter fields. }
\label{fig1}
\end{figure}

In general, the complete set of Feynman graphs corresponding to a given
scattering process includes irreducible diagrams as well as
reducible.\footnote{Here and below in this section, the term "reducible"
is used with respect to the gauge field only.}
It is important, however, that in many cases a subset of diagrams, consisting
of only reducible ones, can be extracted from the complete set, which
contains contributions remaining finite in the limit $m \to \infty,$
$m$ denoting the masses of the scattering particles. In electrodynamics and
Yang-Mills theories, for instance, this is the case for the
spin-$\frac{1}{2}$ particles, the subset containing all diagrams without
internal lines of the scattering particles [see Fig.~\ref{fig1}(a)], but
not for the spin-0 particles, in which case one has also diagrams of the
type shown in Fig.~\ref{fig1}(b). In the case of quantum gravity, furthermore,
things are even more complicated. Besides diagrams of Fig.~1(b), one has
also diagrams pictured in Fig.~\ref{fig1}(c), which do not disappear in the limit
$m \to \infty,$ since $m$ multiplies the vertices of gravitational
interactions of the particles, i.e., turns out to be not only in the
denominators, but also in the numerators of the Feynman integrals.

We see that the definition of potential via scattering amplitudes
is hardly justified in cases when the gauge field -- matter interaction
is nonlinear in the gauge field. The requirement of {\it one}-particle
reducibility, underlying this definition, seems to be adequate only
for {\it linear} interactions.

Definition of the potential through the scattering amplitudes is not the
only possible way to introduce an independent notion of the gauge field.
It {\it is}, however, if one is interested in giving a {\it gauge-independent}
definition, i.e., the one that would give values for the gauge field, which
are independent of the choice of gauge conditions needed to fix gauge
invariance of the theory.\footnote{One also has to require independence of
the choice of a set of dynamical variables in terms of which the theory is
quantized. This last condition is particularly important in the case of
gravity, where one is free to take any tensor density as a dynamical
parametrization of the metric field.} Actually, it was recently proposed
that, in the case of quantum gravity, such a definition can be given beyond
the S-matrix approach through the introduction of classical point particle
moving in the given gravitational field and playing the role of a measuring
device \cite{dalvit}. In particular, it was shown that the one-loop effective
equations of motion of the point-particle (the effective geodesic equation),
calculated in the weak field approximation in the non-relativistic limit,
turn out to be independent of the gauge conditions fixing the general
covariance \cite{dalvit}. Although this result, undoubtedly, is of
considerable importance on its own, it lies out of the main line of our
concern here, since it is based on the introduction of the classical
point-particle into the functional integral "by hands", which certainly
cannot be justified using consistent limiting procedure of transition
from the underlying quantum field theory to the classical theory.
On the other hand, as was shown in Ref.~\cite{kazakov1}, introduction of
the classical {\it field} matter (scalar field) instead of the point-like
still leads to the gauge-dependent values for the gravitational
field.\footnote{It seems that in the case of ordinary Yang-Mills theories,
inclusion of the classical field matter does solve the gauge-dependence
problem, at least in the low-energy limit, see Ref.~\cite{kazakov2}.}

Turning back to the problem of definition of the gravitational
potential through the scattering amplitudes, we see that
since irreducible diagrams to be dropped out do not disappear even in the
limit $m \to \infty,$ {\it validity of the most attractive property of the
potential defined through the scattering amplitudes is jeopardized by
the fact that the equivalence theorem asserting the gauge-independence
of the S-matrix is applicable only to the whole set of diagrams, containing
irreducible as well as reducible Feynman graphs describing given scattering
process \cite{dewitt1,ktut}.} As will be shown below, {\it the gravitational
potential} constructed in Ref.~\cite{donoghue} (i.e., using
only reducible Feynman diagrams) {\it does depend on the gauge}, loosing
thereby any significance as a means for description of particle interactions.

\section{Generating functionals and Slavnov identities}\label{tools}

As in Ref.~\cite{donoghue}, we consider the gravitational scattering
of two scalar particles with masses $m_1,$ $m_2.$ Dynamics of their quantum
fields denoted by $\phi_1,$ $\phi_2,$ respectively, is described by the
action

\begin{eqnarray}&&
S_{\phi} =  \frac{1}{2}{\displaystyle\int} d^4 x \sqrt{-g}(g^{\mu\nu}\partial_{\mu}\phi \partial_{\nu}\phi - m^2 \phi^2),
~~\phi = \phi_{1,2},~~m = m_{1,2},
\nonumber
\end{eqnarray}
\noindent
while the action for the gravitational field\footnote{Our notation
is $R_{\mu\nu} \equiv R^{\alpha}_{~\mu\alpha\nu} =
\partial_{\alpha}\Gamma^{\alpha}_{\mu\nu} - \cdot\cdot\cdot,
~R \equiv R_{\mu\nu} g^{\mu\nu}, ~g\equiv \det g_{\mu\nu},
~g_{\mu\nu} = {\rm sgn}(+,-,-,-).$
Dynamical variables of the gravitational field
$h_{\mu\nu} = g_{\mu\nu} - \eta_{\mu\nu},
\eta_{\mu\nu} = {\rm diag}\{+1,-1,-1,-1\}.$}
\begin{eqnarray}&&
S = - \frac{1}{k^2}{\displaystyle\int} d^4 x \sqrt{-g}R,
\nonumber
\end{eqnarray}
$k$ being the gravitational constant.\footnote{We choose units in
which $c = \hbar = k = 1$ from now on.}

The action $S + S_{\phi_1} + S_{\phi_2}$ is invariant under the following
(infinitesimal) gauge transformations\footnote{Indices of the
functions $F, \xi$, as well as of the ghost fields below,
are raised and lowered, if convenient, with the help of Minkowski metric $\eta_{\mu\nu}$.}
\begin{eqnarray}&&
\delta h_{\mu\nu} = \xi^{\alpha}\partial_{\alpha}h_{\mu\nu}
+ (\eta_{\mu\alpha} + h_{\mu\alpha})\partial_{\nu}\xi^{\alpha}
+ (\eta_{\nu\alpha} + h_{\nu\alpha})\partial_{\mu}\xi^{\alpha}
\nonumber\\&&
\equiv D_{\mu\nu}^{\alpha}(h)\xi_{\alpha},
\nonumber\\&&
~~\delta\phi = \xi^{\alpha}\partial_{\alpha}\phi \equiv
\tilde{D}^{\alpha}(\phi)\xi_{\alpha},
\nonumber
\end{eqnarray}
where $\xi^{\alpha}$ are the (infinitesimal) gauge functions.
The generators $D,\tilde{D}$ span the closed algebra
\begin{eqnarray}&&
D_{\mu\nu}^{\alpha,\sigma\lambda} D_{\sigma\lambda}^{\beta}
- D_{\mu\nu}^{\beta,\sigma\lambda} D_{\sigma\lambda}^{\alpha}
= f_{~~~\gamma}^{\alpha\beta} D_{\mu\nu}^{\gamma},
\nonumber\\&&
\tilde{D}^{\alpha}_{1} \tilde{D}^{\beta}
- \tilde{D}^{\beta}_1 \tilde{D}^{\alpha} = f^{\alpha\beta}_{~~~\gamma} \tilde{D}^{\gamma},
\nonumber
\end{eqnarray}
the "structure constants" $f^{\alpha\beta}_{~~~\gamma}$ being defined by
\begin{eqnarray}&&
f_{~~~\gamma}^{\alpha\beta}\xi_{\alpha}\eta_{\beta} =
\xi_{\alpha}\partial^{\alpha}\eta_{\gamma}
- \eta_{\alpha}\partial^{\alpha}\xi_{\gamma}.
\nonumber
\end{eqnarray}

Let the gauge invariance be fixed by the term
\begin{eqnarray}&&
S_{gf} = \frac{1}{2\xi}\eta^{\alpha\beta} F_{\alpha} F_{\beta},
\nonumber\\&&
~~F_{\alpha} = \partial^{\mu} h_{\mu\alpha} - \frac{1}{2}\partial_{\alpha} h,
~~h \equiv \eta^{\mu\nu} h_{\mu\nu}.
\nonumber
\end{eqnarray}

Next, introducing the Faddeev-Popov ghost fields
$C_{\alpha}, \bar{C}^{\alpha}$ we write the Faddeev-Popov quantum action
\cite{faddeev}
\begin{eqnarray}
S_{FP} = S + S_{\phi_1} + S_{\phi_2} + S_{gf}
+ \bar{C}^{\beta}F_{\beta}^{,\mu\nu}D_{\mu\nu}^{\alpha}C_{\alpha}.
\nonumber
\end{eqnarray}
$S_{FP}$ is still invariant under the following BRST transformations \cite{brst}
\begin{eqnarray}\label{brst}&&
\delta h_{\mu\nu} = D_{\mu\nu}^{\alpha}(h)C_{\alpha}\lambda,
\nonumber\\&&
\delta \phi = \tilde{D}^{\alpha}(\phi)C_{\alpha}\lambda,
\nonumber\\&&
\delta C_{\gamma} = - \frac{1}{2}f^{\alpha\beta}_{~~~\gamma}C_{\alpha}C_{\beta}\lambda,
\nonumber\\&&
\delta \bar{C}^{\alpha} = \frac{1}{\xi}F^{\alpha}\lambda,
\end{eqnarray}
$\lambda$ being a constant anticommuting parameter.

The generating functional of Green functions\footnote{For brevity, the product symbol,
as well as tensor indices of the fields $h_{\mu\nu},
C_{\alpha}, \bar{C}^{\alpha},$ is omitted in the path integral measure.}
\begin{eqnarray}&&
Z[T,J,\bar{\beta},\beta,K,\tilde{K},L]
\nonumber\\&&
= {\displaystyle\int}dh d\phi dC d\bar{C} \exp\{i (\Sigma
+ \bar{\beta}^{\alpha}C_{\alpha} + \bar{C}^{\alpha}\beta_{\alpha} + T^{\mu\nu}h_{\mu\nu} + J\phi )\},
\nonumber
\end{eqnarray}
where $J = \{J_{1,2}\},~~d\phi \equiv d\phi_1 d\phi_2,
~~J\phi \equiv J_1\phi_1 + J_2\phi_2$,
and
\begin{eqnarray}&&
\Sigma = S_{FP}
+ K^{\mu\nu}D_{\mu\nu}^{\alpha}C_{\alpha}
+ \tilde{K}\tilde{D}^{\alpha}C_{\alpha}
+ L^{\gamma} \frac{1}{2} f^{\alpha\beta}_{~~~\gamma}C_{\alpha}C_{\beta},
\nonumber
\end{eqnarray}
$K^{\mu\nu}(x), ~\tilde{K}(x)$ (anticommuting),$ ~L^{\alpha}(x)$(commuting)
being the BRST transformation sources \cite{zinnjustin}.

To determine the dependence of field-theoretical quantities on the gauge
parameter $\xi$, we modify the quantum action adding the term
\begin{eqnarray}
Y F_{\alpha}\bar{C}^{\alpha},
\nonumber
\end{eqnarray}
\noindent
$Y$ being a constant anticommuting parameter \cite{nielsen}.
Thus we write the generating functional of Green functions as
\begin{eqnarray}\label{genernew}&&
Z[T,J,\bar{\beta},\beta,K,\tilde{K},L,Y]
= {\displaystyle\int}dh d\phi dC d\bar{C} \exp\{i (\Sigma
\nonumber\\&&
+ Y F_{\alpha}\bar{C}^{\alpha}
+ \bar{\beta}^{\alpha}C_{\alpha} + \bar{C}^{\alpha}\beta_{\alpha} + T^{\mu\nu}h_{\mu\nu} + J\phi )\}.
\end{eqnarray}

Finally, we introduce the generating functional of
connected Green functions
\begin{eqnarray}
W[T,J,\bar{\beta},\beta,K,\tilde{K},L,Y]=
- i \ln Z[T,J,\bar{\beta},\beta,K,\tilde{K},L,Y],
\nonumber
\end{eqnarray}
and then define the effective action $\Gamma$ in the usual way
as the Legendre transform of $W$ with respect to the mean fields
\begin{eqnarray}&&\label{mean}
h_{\mu\nu} = \frac{\delta W}{\delta T^{\mu\nu}},
~~\phi = \frac{\delta W}{\delta J},
~~C_{\alpha} = \frac{\delta W}{\delta\bar{\beta}^{\alpha}},
~~\bar{C}^{\alpha} = - \frac{\delta W}{\delta\beta_{\alpha}},
\end{eqnarray}
(denoted by the same symbols as the corresponding field operators):
\begin{eqnarray}&&
\Gamma[h,\phi,C,\bar{C},K,\tilde{K},L,Y]
\nonumber\\&&
= W [T,J,\bar{\beta},\beta,K,\tilde{K},L,Y]
-  \bar{\beta}^{\alpha}C_{\alpha} - \bar{C}^{\alpha}\beta_{\alpha} - T^{\mu\nu}h_{\mu\nu} - J\phi.
\nonumber
\end{eqnarray}

Evaluation of derivatives of diagrams with respect to the gauge parameters
is a more easy task than their direct calculation in arbitrary
gauge.\footnote{In actual quantum gravity calculations, this fact was
first used in \cite{tutin} to evaluate divergences of the Einstein gravity
in arbitrary gauge off the mass shell.} This is because these derivatives
can be expressed through another set of diagrams with more simple structure.
The rules for such a transformation of diagrams are conveniently summarized
in the Slavnov identities corresponding to the generating functional
(\ref{genernew}). Since these identities are widely used in what follows,
their derivation will be briefly described below \cite{nielsen}.

First of all, we perform a BRST shift (\ref{brst}) of integration
variables in the path integral (\ref{genernew}). Equating the variation
to zero we obtain the following identity
\begin{eqnarray}&&\label{slav}
{\displaystyle\int}dh d\phi dC d\bar{C}
\left[i Y \bar{C}^{\alpha} F_{\alpha}^{,\mu\nu} D^{\beta}_{\mu\nu} C_{\beta}
+ i \frac{Y}{\xi} F_{\alpha}^{2}
+ T^{\mu\nu} \frac{\delta}{\delta K^{\mu\nu}}
+ J \frac{\delta}{\delta\tilde{K}}
- \bar{\beta}^{\alpha}\frac{\delta}{\delta L^{\alpha}}
- i \beta_{\alpha}\frac{F^{\alpha}}{\xi}
\right]
\nonumber\\&&
\times\exp\{i (\Sigma
+ Y F_{\alpha}\bar{C}^{\alpha}
+ \bar{\beta}^{\alpha}C_{\alpha} + \bar{C}^{\alpha}\beta_{\alpha}
+ T^{\mu\nu}h_{\mu\nu} + J\phi)\} = 0.
\end{eqnarray}
\noindent
Next, the second term in the square brackets in Eq.~(\ref{slav})
can be transformed with the help of the quantum ghost equation of motion,
obtained by performing a shift $\bar{C} \to \bar{C} + \delta\bar{C}$
of integration variables in the functional integral (\ref{genernew}):
\begin{eqnarray}&&
{\displaystyle\int}dh d\phi dC d\bar{C}
\left[F_{\gamma}^{,\mu\nu}D_{\mu\nu}^{\alpha}C_{\alpha}
- Y F_{\gamma} + \beta_{\gamma} \right]
\exp\{\cdot\cdot\cdot\} = 0,
\nonumber
\end{eqnarray}
\noindent
from which it follows that
\begin{eqnarray}&&
Y {\displaystyle\int}dh d\phi dC d\bar{C}
\left[i \bar{C}^{\gamma}F_{\gamma}^{,\mu\nu}D_{\mu\nu}^{\alpha}C_{\alpha}
+ \beta_{\gamma}\frac{\delta}{\delta\beta_{\gamma}}\right]
\exp\{\cdot\cdot\cdot\} = 0,
\nonumber
\end{eqnarray}
\noindent
where we used the property $Y^2 = 0$, and omitted the expression
$\delta\beta_{\gamma}/\delta\beta_{\gamma} \sim \delta(0)$.
Putting this all together, we rewrite Eq.~(\ref{slav})
\begin{eqnarray}
\left( T^{\mu\nu}\frac{\delta}{\delta K^{\mu\nu}}
+ J\frac{\delta}{\delta \tilde{K}}
- \bar{\beta}^{\alpha}\frac{\delta}{\delta L^{\alpha}}
- \frac{1}{\xi} \beta_{\alpha}F^{\alpha,\mu\nu}\frac{\delta}{\delta T^{\mu\nu}}
- Y \beta_{\gamma}\frac{\delta}{\delta\beta_{\gamma}}
- 2 Y\xi\frac{\partial}{\partial\xi}
\right) Z  = 0.
\nonumber
\end{eqnarray}
This is the Slavnov identity for the generating functional of Green functions
we are looking for.
In terms of the generating functional of connected Green functions,
it looks like
\begin{eqnarray}\label{slav2}
T^{\mu\nu}\frac{\delta W}{\delta K^{\mu\nu}}
+ J\frac{\delta W}{\delta \tilde{K}}
- \bar{\beta}^{\alpha}\frac{\delta W}{\delta L^{\alpha}}
- \frac{1}{\xi} \beta_{\alpha}F^{\alpha,\mu\nu}\frac{\delta W}{\delta T^{\mu\nu}}
- Y \beta_{\gamma}\frac{\delta W}{\delta\beta_{\gamma}}
- 2 Y\xi\frac{\partial W}{\partial\xi} = 0.
\end{eqnarray}
\noindent
It can be transformed further into an identity for the generating functional
of proper vertices: with the help of equations
\begin{eqnarray}&&\label{meaninv}
T^{\mu\nu} =  - \frac{\delta \Gamma}{\delta h_{\mu\nu}},
~~J =  - \frac{\delta \Gamma}{\delta \phi},
~~\bar{\beta}^{\alpha} = \frac{\delta \Gamma}{\delta C_{\alpha}},
~~\beta_{\alpha} = - \frac{\delta \Gamma}{\delta\bar{C}^{\alpha}},
\end{eqnarray}
which are the inverse of Eqs.~(\ref{mean}),
and the relations
\begin{eqnarray}
\frac{\delta W}{\delta K^{\mu\nu}} = \frac{\delta \Gamma}{\delta K^{\mu\nu}},
~~\frac{\delta W}{\delta \xi} = \frac{\delta \Gamma}{\delta \xi}, {\rm ~~etc.}
\nonumber
\end{eqnarray}
we rewrite Eq.~(\ref{slav2})
\begin{eqnarray}&&
\frac{\delta \Gamma}{\delta h_{\mu\nu}}\frac{\delta \Gamma}{\delta K^{\mu\nu}}
+ \frac{\delta \Gamma}{\delta \phi}\frac{\delta \Gamma}{\delta \tilde{K}}
+ \frac{\delta \Gamma}{\delta C_{\alpha}}\frac{\delta \Gamma}{\delta L^{\alpha}}
- \frac{F^{\alpha}}{\xi} \frac{\delta \Gamma}{\delta\bar{C}^{\alpha}}
+ Y \frac{\delta \Gamma}{\delta\bar{C}^{\alpha}}\bar{C}^{\alpha}
+ 2 Y\xi\frac{\partial \Gamma}{\partial\xi} = 0.
\nonumber
\end{eqnarray}
\noindent
Written down via the reduced functional
\begin{eqnarray}&&
{\Gamma_0} = \Gamma - \frac{1}{2\xi}F_{\alpha} F^{\alpha}
- Y F_{\sigma}\bar{C}^{\sigma},
\nonumber
\end{eqnarray}
the latter equation takes particularly simple form
\begin{eqnarray}\label{slav4}&&
\frac{\delta\Gamma_0}{\delta h_{\mu\nu}}\frac{\delta\Gamma_0}{\delta K^{\mu\nu}}
+ \frac{\delta\Gamma_0}{\delta \phi}\frac{\delta\Gamma_0}{\delta \tilde{K}}
+ \frac{\delta\Gamma_0}{\delta C_{\sigma}}\frac{\delta\Gamma_0}{\delta L^{\sigma}}
+ 2 Y \xi\frac{\partial\Gamma_0}{\partial\xi} = 0.
\end{eqnarray}

\section{Gauge dependence of the one-particle-reducible gravitational
potential}\label{loop}

Let us now turn to the explicit evaluation of the $\xi$-dependence of the
one-loop contribution to the potential. Its general structure is
shown in Fig.~\ref{fig1}(a). In view of the assumed reducibility,
corrections to the vertices and graviton propagator, which are the building
blocks for the potential, can be considered separately.
Let us note first of all that the (tree) graviton propagators, with respect
to which the potential is reducible, can be considered gauge-independent.
Indeed, at the one-loop level, each of these propagators has one of its
ends attached to the tree $\phi-h-\phi$-vertex with the $\phi$-lines
on the mass shell. This combination is gauge-independent on the same grounds
as is the $S$-matrix at the tree level. Thus, we have to consider
only the proper $\phi-h-\phi$-vertex and the graviton self-energy.
To evaluate the $\xi$-derivative of these quantities, we use the Slavnov
identity (\ref{slav4}). Extracting terms proportional to the source $Y,$
we get
\begin{eqnarray}\label{slav4y}&&
2\xi\frac{\partial\Gamma_{1}}{\partial\xi} =
 \frac{\delta\Gamma_{1}}{\delta h_{\mu\nu}}\frac{\delta\Gamma_{2}}{\delta K^{\mu\nu}}
+ \frac{\delta\Gamma_{1}}{\delta \phi}\frac{\delta\Gamma_{2}}{\delta \tilde{K}},
\end{eqnarray}
\noindent
where $\Gamma_{1,2}$ are defined by
$$\Gamma_{1} = \Gamma_{0}|_{Y=0}, ~~\Gamma_{2}
= \frac{\partial\Gamma_{0}}{\partial Y}.$$

At the one-loop level, Eq.~(\ref{slav4y}) is just\footnote{Enclosed
in the round brackets is the number of loops in a diagram representing given
term.}
\begin{eqnarray}\label{slav4y1}&&
2\xi\frac{\partial\Gamma^{(1)}_{1}}{\partial\xi} =
\frac{\delta \Gamma^{(0)}_{1}}{\delta h_{\mu\nu}}\frac{\delta\Gamma^{(1)}_{2}}{\delta K^{\mu\nu}},
\end{eqnarray}
since the external scalar lines are on the mass shell
$$\frac{\delta S_{\phi}}{\delta \phi} = 0.$$
\noindent

\begin{figure}
\epsfxsize=15,5cm\epsfbox{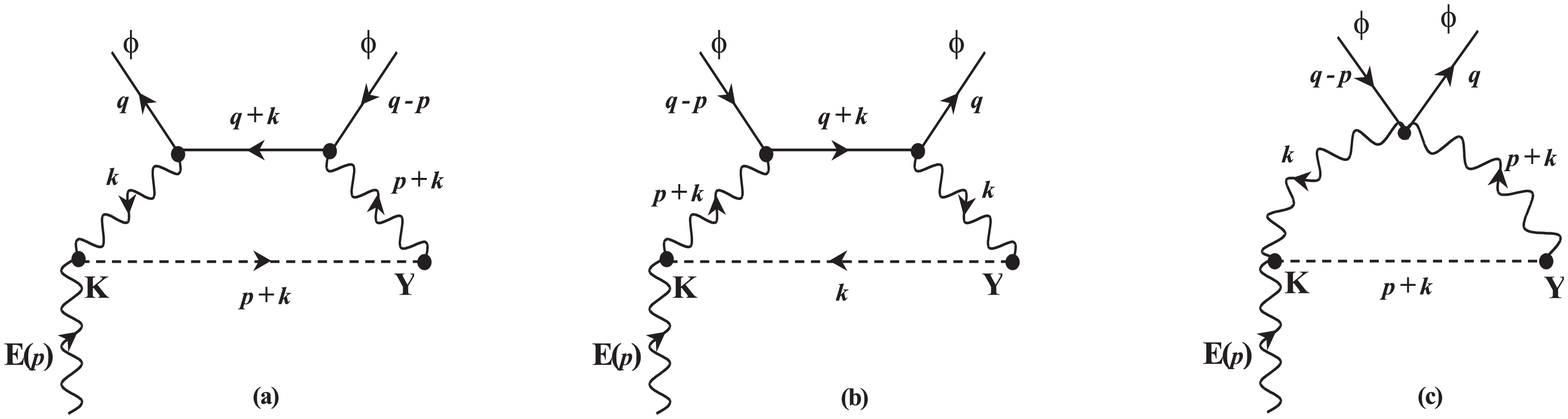}
\vspace{1cm}
\caption{Diagrams with two scalar and one graviton external lines,
responsible for the non-vanishing of the $\xi$-dependent contribution
to the one-particle-reducible gravitational potential.
Solid lines represent scalar particles, dashed lines ghosts.}
\label{fig2}
\end{figure}

\begin{figure}
\epsfxsize=15,5cm\epsfbox{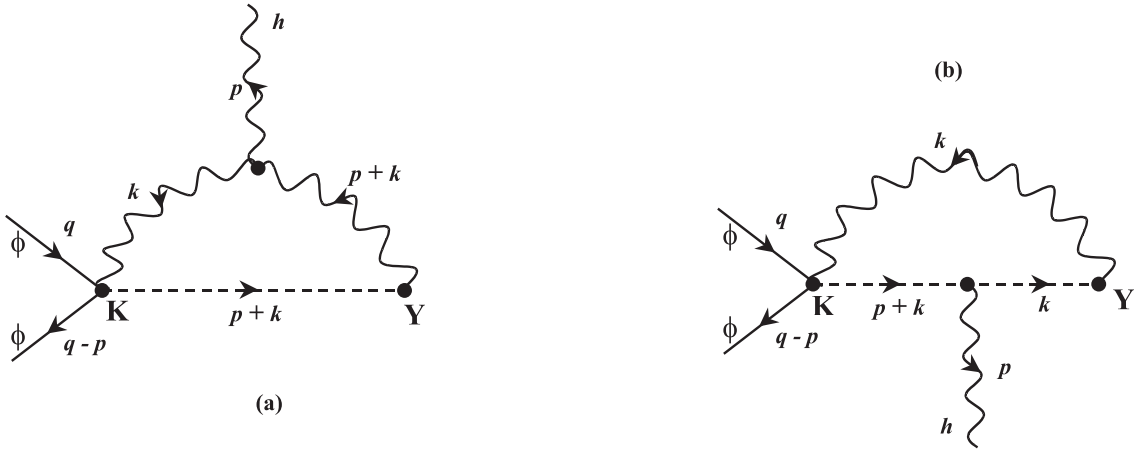}
\vspace{1cm}
\caption{Diagrams representing the part of the $\xi$-dependent contribution
to the gravitational form factors of scalar particles, that cancels
the corresponding contribution coming from the graviton self-energy
(see Fig.~4) in the course of construction of the potential.}
\label{fig3}
\end{figure}

Graphs representing the $\xi$-derivatives of the form factors according
to the right hand side of Eq.~(\ref{slav4y1}), are shown in
Figs.~\ref{fig2},~\ref{fig3}.

\begin{figure}
\epsfxsize=15,5cm\epsfbox{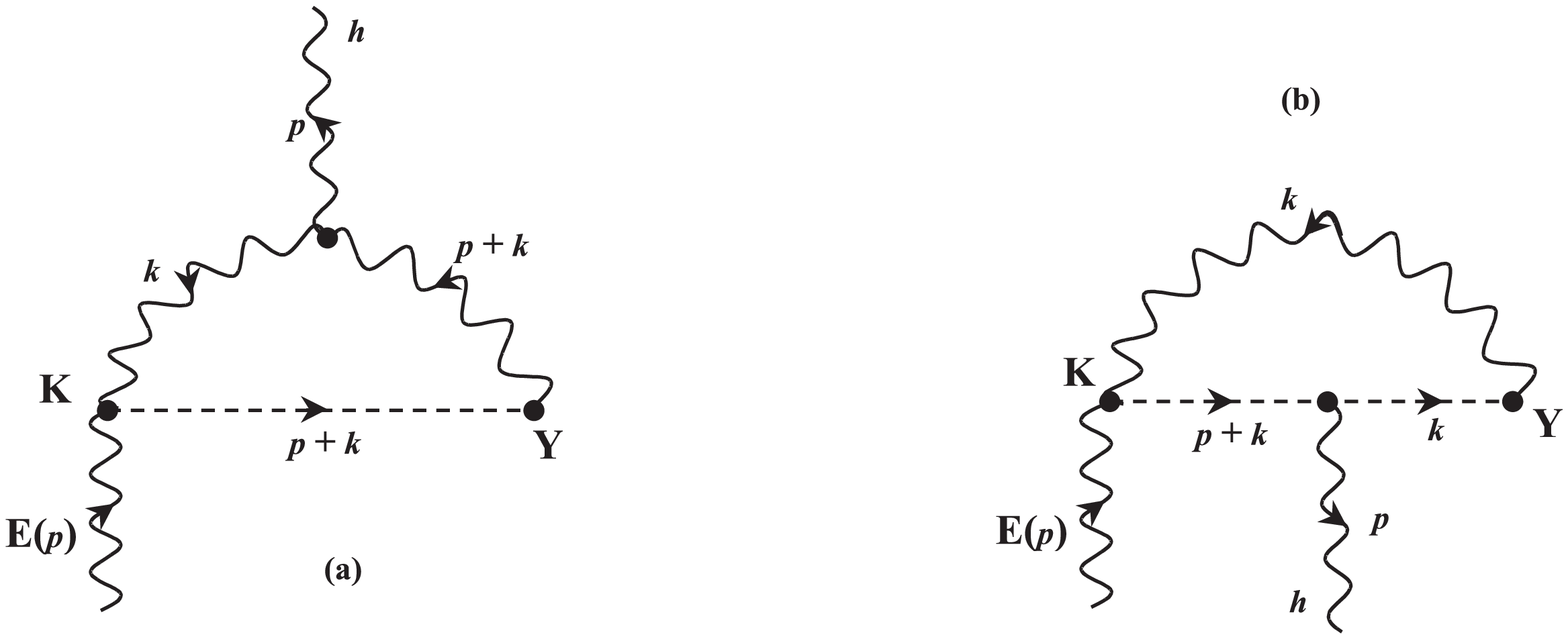}
\vspace{1cm}
\caption{Diagrams representing the $\xi$-dependent contribution
to the graviton self-energy.}
\label{fig4}
\end{figure}

Diagrams of Fig.~\ref{fig3} need not be calculated explicitly.
It is easy to see that they just cancel the $\xi$-dependent contribution
to the graviton self-energy when the potential is being constructed.
Indeed, according to Eq.~(\ref{slav4y1}), this contribution is given
by the diagrams of Fig.~\ref{fig4}. In the course of construction of the
potential, the two $h$-lines of the graviton self-energy are connected
to the $\phi-h-\phi$-vertices by the graviton propagators. When these
propagators are attached to the left most vertices in Figs.~\ref{fig4}(a),(b),
we get exactly the diagrams of Figs.~\ref{fig3}(a),(b), respectively,
but with the opposite sign, because it follows from
Eqs.~(\ref{mean}),(\ref{meaninv}) that
\begin{eqnarray}&&
\frac{\delta^2 S}{\delta h_{\mu\nu} \delta h_{\alpha\beta}}
\frac{\delta^2 W^{(0)}}{\delta T^{\alpha\beta} \delta T^{\gamma\delta}}
= - \delta^{\mu\nu}_{\gamma\delta}.
\nonumber
\end{eqnarray}

Thus, explicit calculation of diagrams of Fig.~\ref{fig2} is needed only.
Their analytic expressions
\begin{eqnarray}&&\label{int}
I_{2(a)}(p,q) =
\frac{-i E^{\mu\nu}(p)}{2\sqrt{\varepsilon_{q}\varepsilon_{q-p}}}
~\mu^{\epsilon} {\displaystyle\int} \frac{d^{4-\epsilon} k}{(2\pi)^4}
\left\{
\frac{1}{2} W^{\alpha\beta\gamma\delta} q_{\gamma}(k_{\delta} + q_{\delta})
- m^2 \frac{\eta^{\alpha\beta}}{2}
\right\} G_{\phi}
\nonumber\\&&
\times\left\{
\frac{1}{2} W^{\rho\tau\sigma\lambda}(q_{\sigma} - p_{\sigma} )
(k_{\lambda} + q_{\lambda}) - m^2 \frac{\eta^{\rho\tau}}{2}
\right\}
\xi D^{(0)\eta}_{\rho\tau}\tilde{G}_{\eta}^{\xi}(k+p),
\nonumber\\&&
\times\tilde{G}_{\xi}^{\zeta}(p+k)
\left\{k_{\zeta} \delta_{\mu\nu}^{\chi\theta} -
\delta_{\zeta\mu}^{\chi\theta} (k_{\nu} + p_{\nu}) -
\delta_{\zeta\nu}^{\chi\theta} (k_{\mu} + p_{\mu})\right\}
G_{\chi\theta\alpha\beta}(k),
\end{eqnarray}
\begin{eqnarray}&&
I_{2(b)}(p,q) = I_{2(a)}(p,p-q),
\nonumber
\end{eqnarray}
\begin{eqnarray}&&\label{intc}
I_{2(c)}(p,q) =
\frac{-i E^{\mu\nu}(p)}{2\sqrt{\varepsilon_{q}\varepsilon_{q-p}}}
~\mu^{\epsilon} {\displaystyle\int} \frac{d^{4-\epsilon} k}{(2\pi)^4}
\left\{\left[
- \frac{1}{2}(\delta^{\sigma\lambda}_{\zeta\xi} \eta^{\tau\rho} +
\delta^{\tau\rho}_{\zeta\xi} \eta^{\sigma\lambda})
\right.
\right.
\nonumber\\&&
\left.
\left.
+ (\delta_{\zeta\omega}^{\tau\rho}\delta_{\xi\omega}^{\sigma\lambda}
+ \delta_{\xi\omega}^{\tau\rho}\delta_{\zeta\omega}^{\sigma\lambda})
+ \frac{\eta_{\zeta\xi}}{4} W^{\tau\rho\sigma\lambda}
\right]
q^{\xi}(q^{\zeta}-p^{\zeta})
- \frac{m^2}{4} W^{\tau\rho\sigma\lambda}
\right\}G_{\tau\rho\chi\theta}(k)
\nonumber\\&&
\times \left\{k_{\alpha} \delta_{\mu\nu}^{\chi\theta} -
\delta_{\mu\alpha}^{\chi\theta} (k_{\nu} + p_{\nu}) -
\delta_{\nu\alpha}^{\chi\theta} (k_{\mu} + p_{\mu})\right\}
\tilde{G}^{\alpha}_{\beta}(p+k)
\xi D^{(0)\gamma}_{\sigma\lambda}\tilde{G}_{\gamma}^{\beta}(k+p),
\end{eqnarray}
\noindent
where the following notation is introduced:
\begin{eqnarray}
W^{\alpha\beta\gamma\delta} =
\eta^{\alpha\beta} \eta^{\gamma\delta}
- \eta^{\alpha\gamma} \eta^{\beta\delta}
- \eta^{\alpha\delta} \eta^{\beta\gamma},
~~\delta_{\alpha\beta}^{\mu\nu} = \frac{1}{2}(\delta_{\alpha}^{\mu}\delta_{\beta}^{\nu}
+ \delta_{\alpha}^{\nu}\delta_{\beta}^{\mu}),
\nonumber
\end{eqnarray}
\noindent
$G_{\mu\nu\sigma\lambda}$ is the graviton propagator defined by
$$\frac{\delta^2 S}{\delta h_{\rho\tau}\delta h_{\mu\nu}}
G_{\mu\nu\sigma\lambda} = - \delta_{\sigma\lambda}^{\rho\tau},$$
\begin{eqnarray}&&
G_{\mu\nu\sigma\lambda} =
- W_{\mu\nu\sigma\lambda}\frac{1}{\Box}
+ (\xi - 1) (\eta_{\mu\sigma} \partial_{\nu} \partial_{\lambda}
+ \eta_{\mu\lambda} \partial_{\nu} \partial_{\sigma}
+ \eta_{\nu\sigma} \partial_{\mu} \partial_{\lambda}
+ \eta_{\nu\lambda} \partial_{\mu} \partial_{\sigma}) \frac{1}{\Box^2},
\nonumber
\end{eqnarray}
\noindent
$\tilde{G}^{\alpha}_{\beta}$ is the ghost propagator
\begin{eqnarray}&&
\tilde{G}^{\alpha}_{\beta} = - \frac{\delta^{\alpha}_{\beta}}{\Box},
\nonumber
\end{eqnarray}
satisfying
\begin{eqnarray}&&
F_{\alpha}^{,\mu\nu}D^{(0)\beta}_{\mu\nu}\tilde{G}^{\gamma}_{\beta}
= - \delta_{\alpha}^{\gamma},
~~D^{(0)\alpha}_{\mu\nu} \equiv D^{\alpha}_{\mu\nu}(h=0),
\nonumber
\end{eqnarray}
\noindent
$G_{\phi}$ is the scalar particle propagator
$$G_{\phi} =  \frac{1}{\Box + m^2},$$
\noindent
$E^{\mu\nu}$ stands for the linearized Einstein tensor
\begin{eqnarray}
E^{\mu\nu} = R^{\mu\nu} - \frac{1}{2}\eta^{\mu\nu} R_{\alpha\beta} \eta^{\alpha\beta},
~~R_{\mu\nu} = \frac{1}{2}(\partial^{\alpha}\partial_{\mu} h_{\alpha\nu}
+ \partial^{\alpha}\partial_{\nu} h_{\alpha\mu}
- \Box h_{\mu\nu} - \partial_{\mu}\partial_{\nu} h),
\nonumber
\end{eqnarray}
\noindent
$\mu$ -- arbitrary mass scale, $\varepsilon_q = \sqrt{q^2 + m^2},$
and $\epsilon = 4 - d,$ $d$ being the dimensionality of space-time.
\noindent
To simplify the tensor structure of diagrams Fig.~\ref{fig2},
the use has been made of the identity
\begin{eqnarray}&&
\frac{1}{\xi}F^{\alpha,\mu\nu} G_{\mu\nu\sigma\lambda}(x) =
- D^{(0)\beta}_{\sigma\lambda}\tilde{G}_{\beta}^{\alpha}(x),
\nonumber
\end{eqnarray}
which is nothing but the well-known {\it first} Slavnov identity at the
tree level; it is easily obtained differentiating Eq.~(\ref{slav2}) twice
with respect to $\beta_{\alpha}$ and $T^{\mu\nu}$, and setting all the
sources equal to zero.

Let us begin with evaluation of the diagram of Fig.~\ref{fig2}(a).
This takes most of efforts.

The tensor multiplication in Eq.~(\ref{int}) is conveniently performed
with the help of the new tensor package for the REDUCE system \cite{reduce}
\begin{eqnarray}&&\label{int1}
I_{2(a)}(p,q)
= \frac{- i E^{\mu\nu}(p)}{2\sqrt{\varepsilon_{q}\varepsilon_{q-p}}}
~\mu^{\epsilon} {\displaystyle\int}
\frac{d^{4-\epsilon} k}{(2\pi)^4} \frac{1}{k^4}\frac{1}{(k+p)^4}
\frac{1}{m^2-(k+q)^2}
\nonumber\\&&
\times \xi [
\eta_{\mu\nu} k^2 m^2 \{(kp) - (kq)\}\{k^2 + 2 (kq)\}
+ k_{\mu} k_{\nu} k^4 \xi (p^2 - 2 m^2)
\nonumber\\&&
+ 2 k_{\mu} k_{\nu} k^2 (kq) (2 \xi - 1) (p^2 - 2 m^2)
+ 4 k_{\mu} (k_{\nu} + p_{\nu}) (kq)^2 (\xi - 1) (p^2 - 2 m^2)
\nonumber\\&&
+ k_{\mu} p_{\nu} k^4 (- 2 \xi m^2 + \xi p^2 - 2 m^2)
+ 2 k_{\mu} p_{\nu} k^2 (kq)(- 4 \xi m^2 + 2 \xi p^2 - p^2)
\nonumber\\&&
+  2 k_{\mu} q_{\nu} k^4 (p^2 - m^2)
+  4 k_{\mu} q_{\nu} k^2 (kq) (p^2 - m^2)
- 2 p_{\mu} p_{\nu} k^2 m^2 \{k^2 + 2 (kq)\}
\nonumber\\&&
+ 2 p_{\mu} q_{\nu} k^2 \xi \{(kp) - (kq)\} \{k^2 + 4 (kq)\}
+ 4 p_{\mu} q_{\nu} k^2 (kq) \{ p^2 - m^2 + (kq) - (kp)\}
\nonumber\\&&
+ 8 p_{\mu} q_{\nu} (kq)^2 (\xi-1)\{(kp) - (kq)\}
+ 2 p_{\mu} q_{\nu} k^4 (p^2 - m^2)
\nonumber\\&&
+ 2 q_{\mu} q_{\nu} k^4 \{(kq)- (kp)\}
+ 4 q_{\mu} q_{\nu} k^2 (kq) \{(kq) - (kp)\}
]
\end{eqnarray}
\noindent
Evaluation of the loop integrals can be automatized to a considerable
extent if the Schwinger parametrization of denominators
in Eq.~(\ref{int1}) is used
\begin{eqnarray}&&
\frac{1}{k^4} = \int_{0}^{\infty} dy~y \exp\{y k^2\},
~~\frac{1}{(k + p)^4} = \int_{0}^{\infty} dx~x \exp\{x (k + p)^2\},
\nonumber\\&&
\frac{1}{k^2 + 2 (kq)} = - \int_{0}^{\infty} dz \exp\{z [k^2 + 2 (kq)]\}.
\nonumber
\end{eqnarray}
It is convenient to apply these formulae as they stand, i.e., eluding
cancellation of the $k^2$ factors in Eq.~(\ref{int1}).
The $k$-integrals are then evaluated using
\begin{eqnarray}&&
\int~d^{d}k \exp\{ k^2 (x + y + z) + 2 k^{\mu} (x p_{\mu} + z q_{\mu})\}
\nonumber\\&&
= i \left(\frac{\pi}{x + y + z}\right)^{\frac{d}{2}}
\exp\left\{\frac{p^2 x y - m^2 z^2}{x + y + z}\right\},
\nonumber
\end{eqnarray}
\begin{eqnarray}&&
\int~d^{d}k ~k_{\alpha}
\exp\{ k^2 (x + y + z) + 2 k^{\mu} (x p_{\mu} + z q_{\mu})\} =
\nonumber\\&&
= i \left(\frac{\pi}{x + y + z}\right)^{\frac{d}{2}}
\exp\left\{\frac{p^2 x y - m^2 z^2}{x + y + z}\right\}
\left[- \frac{x p_{\alpha} + z q_{\alpha}}{x + y + z}\right],
\nonumber
\end{eqnarray}
\noindent
etc. up to six $k$-factors in the integrand.

From now on, all formulae will be written out for the sum
$$\tilde{I}_2 \equiv I_{2(a)}(p,q) + I_{2(b)}(p,q).$$

Changing the integration variables $(x,y,z)$ to $(t,u,v)$ via
$$x = \frac{t (1 + t + u) v^2}{m^2 (1 + \alpha t u)},
~~y = \frac{u (1 + t + u) v^2}{m^2 (1 + \alpha t u)},
~~z = \frac{ (1 + t + u) v^2}{m^2 (1 + \alpha t u)},
~~\alpha \equiv - \frac{p^2}{m^2},$$
integrating $v$ out, subtracting the ultraviolet divergence\footnote{
Since we are interested only in the non-analytic at $p^2=0$ terms
responsible for the long-range quantum corrections, particularities of
the subtraction scheme are immaterial.}
$$\tilde{I}^{{\rm div}} = \frac{1}{32\pi^2 \epsilon}
\left(\frac{\mu}{m}\right)^{\epsilon} E^{\mu\nu}(p)\eta_{\mu\nu}
\xi^2 (p^2 - 2 m^2),$$ setting $\epsilon = 0$,
and retaining only the terms giving rise to the roots and logarithms of
$p^2/m^2,$ leading at $p^2 \to 0$, we obtain
\begin{eqnarray}&&\label{int2}
(\tilde{I}_2 - \tilde{I}_2^{{\rm div}})_{{\epsilon} \to 0}
\nonumber\\&&
=  \frac{E^{\mu\nu}(p)\xi}{32\pi^2 \sqrt{\varepsilon_{q}\varepsilon_{q-p}}}
\int_{0}^{\infty}\int_{0}^{\infty} du dt \left\{
\frac{8 m^2 (\xi - 1)}{p^2 D N^3}
\left(q_{\mu} q_{\nu} - \frac{m^2}{p^2} p_{\mu} p_{\nu} \right)
\left(6 - \frac{9}{D} + \frac{4}{D^2} \right)
\right.
\nonumber\\&&
\left.
+ \frac{\eta_{\mu\nu} m^2}{D N}
\left( 1 - \frac{5}{D} + \frac{4}{D^2} \right)
+ \frac{4 \eta_{\mu\nu} m^4 \xi}{D N^3 p^2}\left(3 - \frac{2}{D} \right)
\right.
\nonumber\\&&
\left.
+ \frac{8\xi m^2}{D N^{2}}
\left[ \frac{p_{\mu} p_{\nu}}{p^2}
\left(\xi - \frac{\xi}{D} - \frac{1}{D} + \frac{1}{D^2}\right)
+ \frac{p_{\mu} q_{\nu}}{p^2}
\left(-3 \xi + 1 + \frac{7 \xi}{D} - \frac{3}{D}
- \frac{4 \xi}{D^2} + \frac{2}{D^2}\right)\right]
\right\},
\nonumber\\&&
D \equiv 1 + \alpha u t, ~~N \equiv 1 + u + t.
\end{eqnarray}
Eq.~(\ref{int2}) is written out in such a form that the leading roots
come from the first two lines only. The remaining $(u,t)$-integrals are
evaluated in the Appendix. Using Eqs.~(\ref{roots}) one readily sees that
the terms proportional to $\sqrt{-p^2}$ in Eq.~(\ref{int2}) cancel.
As explained elsewhere (see Ref.~\cite{kazakov3}), this fact allows one
to give a physical interpretation to the root contributions to the form
factors directly in the framework of the effective action method, as
describing quantum deviations of the space-time metric from classical
solutions of the Einstein equations.

It is easy to see also that the diagrams of Fig.~\ref{fig2} are the only
that give rise to the root singularities in the potential defined according
to Ref.~\cite{donoghue}, so the found cancellation proves the
gauge-independence of the $1/r^2$-terms in this potential as well
($r$ being the distance from the source-particle). Let us, therefore,
push our calculations further and turn to the $1/r^3$-terms, i.e., to the
leading logarithms. With the help of Eqs.~(\ref{logs}) of the Appendix,
we get from Eq.~(\ref{int2})
\begin{eqnarray}&&\label{int3}
(\tilde{I}_2 - \tilde{I}_2^{{\rm div}})^{{\rm log}}_{{\epsilon} \to 0}
\equiv \tilde{I}^{{\rm ren}}_{2}
=  - \ln\alpha\frac{E^{\mu\nu}(p)\eta_{\mu\nu}
m^2 \xi^2}{32\pi^2 \sqrt{\varepsilon_{q}\varepsilon_{q-p}}}.
\end{eqnarray}

It remains only to calculate the diagram of Fig.~\ref{fig2}(c).
This is a much easier task than the above calculation, since the loop
does not contain scalar lines.
On dimensional grounds, $I_{2(c)}(p,q)$ has the following structure
\begin{eqnarray}\label{int5}&&
I_{2(c)}(p,q) = \frac{E^{\mu\nu}(p) P_{\mu\nu}(p,q)}
{\sqrt{\varepsilon_{q}\varepsilon_{q-p}}}
\left(\frac{\mu^2}{-p^2}\right)^{\epsilon/2}
\left[\frac{1}{\epsilon} + c\right]
\nonumber\\&&
= \frac{E^{\mu\nu}(p) P_{\mu\nu}(p,q)}
{\sqrt{\varepsilon_{q}\varepsilon_{q-p}}}
\left[\frac{1}{\epsilon} - \frac{1}{2}\ln\left(\frac{-p^2}{\mu^2}\right)
+ c + O(\epsilon)\right],
\end{eqnarray}
where $c$ is some number, and $P_{\mu\nu}(p,q)$ polynomials
in $p_{\mu},q_{\mu}.$
It follows from Eq.~(\ref{int5}) that one can obtain the
logarithmic contribution from divergent one substituting
\begin{eqnarray}&&
\frac{1}{\epsilon} \to - \frac{1}{2}\ln\left(\frac{-p^2}{\mu^2}\right).
\nonumber
\end{eqnarray}
\noindent
$I_{2(c)}$ is ultraviolet divergent. It is important, on the other hand,
that it is free of infrared divergences. Indeed, the integrand
in Eq.~(\ref{intc}) is the sum of products of powers $(p + k)^{n}$
and $k^{l},$ times a polynomial in $p_{\mu},q_{\mu}$.
Since the diagram is logarithmically divergent, we have
$n + l = - 4.$ On the other hand, infrared divergences appear only
if $n \le - 4,$ or $l \le - 4$, and, therefore, we have $l \ge 0$, or
$n \ge 0$. In either case the dimensionally regularized loop integrals
turn into zero.

Now, the calculation is straightforward. To find the ultraviolet divergences,
one sets $p + k \to k$ in the propagators and the vertex factors
(since the degree of divergence is zero), averages over
angles (in $k$-space), and retains only $1/k^4$-terms in the integrand,
changing them to $ 2\pi^2 i/\epsilon$ afterwards. The tensor multiplication
as well as integration over angles in the momentum space is again
performed with the help of the tensor package of Ref.~\cite{reduce}.
Subtracting the $1/\epsilon$ divergence and setting $\epsilon = 0$,
one obtains the following result
\begin{eqnarray}&&
\left.I^{{\rm ren}}_{2(c)}(p,q)\right|_{\rm log}
= \frac{\xi\ln\alpha}{96\pi^2 \sqrt{\varepsilon_{q}\varepsilon_{q-p}}}
E^{\mu\nu}(p) \{\eta_{\mu\nu} m^2 ( - 5  \xi + 2)
- q_{\mu} q_{\nu} ( 4 \xi + 8 ) \}.
\nonumber
\end{eqnarray}
\noindent
The total logarithmic contribution of diagrams of Fig.~\ref{fig2} is
\begin{eqnarray}&&\label{sum}
I^{{\rm ren}}_{2} \equiv
I^{{\rm ren}}_{2(c)}(p,q) + \tilde{I}^{{\rm ren}}_{2}
\nonumber\\&&
= - \frac{\xi\ln\alpha}{48\pi^2 \sqrt{\varepsilon_{q}\varepsilon_{q-p}}}
E^{\mu\nu}(p) \{\eta_{\mu\nu} m^2 ( 4 \xi - 1)
+ q_{\mu} q_{\nu} ( 2 \xi + 4 ) \}.
\end{eqnarray}

Finally, multiplying Eq.~(\ref{sum}) with $m = m_1$ by the graviton
propagator and the tree vertex factor corresponding to the second
particle with $m = m_2,$ and adding the result of this calculation
with $m_1, m_2$ interchanged (and $p \to - p$), we have for the
$\xi$-derivative of the one-loop contribution to the
one-particle-reducible part of the two-particle scattering amplitude,
in the case $|{\bf q}_1| \ll m_1,$ $|{\bf q}_2| \ll m_2,$
\begin{eqnarray}&&\label{main}
\frac{\partial A_{1{\rm PR}}^{(1)}}{\partial\xi}
= - \ln(- p^2)\frac{m_{1} m_{2} (2\xi + 1)}{64\pi^2}.
\end{eqnarray}
\noindent
This completes exposition of the main result of the work.

\section{Conclusion}\label{conclude}

The $1/r^3$ terms in the one-particle-reducible gravitational potential
are thus shown to be $\xi$-dependent, the form of this dependence being
given by the Fourier transform of Eq.~(\ref{main}).
The formal reason for the occurrence
of gauge-dependence should be clear from the considerations of
Sec.~\ref{loop}. The gauge invariance of the classical action
is crucial for the proof of the gauge-independence of the
$S$-matrix \cite{dewitt1,ktut}. Being inhomogeneous in the field
$h_{\mu\nu},$ the generators of the gauge transformations mix vertices
with different number of $h$-lines. The gauge invariance of the scattering
amplitude is therefore preserved only if every combination of vertices,
contributing at a given loop order, is taken into account.
Omission of the irreducible part of the two-particle
scattering amplitude inevitably violates the latter condition, the result
being only the partial cancellation of the gauge-dependent contributions,
found in Sec.~\ref{loop}.

Thus, the one-particle-reducible gravitational potential is irrelevant
to the issue of interpretation of the quantum corrections
to the classical metric.

\section*{Acknowledgments}
I would like to thank Drs. P.I.Pronin and K.V.Stepanyantz
(Department of Theoretical Physics, Moscow State University)
for the help in seizing the working facilities of their powerful
tensor package.

\vspace{1cm}
{\Large \bf Appendix}
\vspace{1cm}

\begin{figure}
\hspace*{3cm}
\epsfxsize=10cm\epsfbox{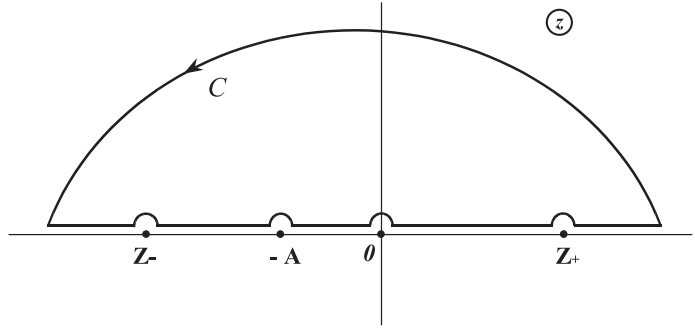}
\vspace{1cm}
\caption{Contour of integration in Eq.~(18)}
\label{fig5}
\end{figure}

The integrals
\begin{eqnarray}&&
J_{nm} \equiv \int_{0}^{\infty}\int_{0}^{\infty}
\frac{du dt}{(A + t + u)^n (B + \alpha t u)^m},
\nonumber
\end{eqnarray}
encountered in Sec.~\ref{loop}, can be evaluated as follows.
\noindent
Consider the auxiliary quantity
\begin{eqnarray}&&
J(A,B) = \int_{0}^{\infty}\int_{0}^{\infty} \frac{du dt}{(A + t + u) (B + \alpha t u)},
\nonumber
\end{eqnarray}
\noindent
where $A,B>0$ are some numbers eventually set equal to 1.
Performing an elementary integration over $u,$ we get
\begin{eqnarray}&&
J(A,B) = \int_{0}^{\infty} dt
~\frac{\ln B - \ln \{\alpha t (A + t)\}}{B - \alpha t (A + t)}.
\nonumber
\end{eqnarray}
\noindent
Now consider the integral
\begin{eqnarray}&&
\tilde{J}(A,B) = \int_{C} dz f(z,A,B),
~~f(z,A,B) = \frac{\ln B - \ln \{\alpha z (A + z)\}}{B - \alpha z (A + z)},
\end{eqnarray}
\noindent
taken over the contour $C$ shown in Fig.~\ref{fig5}.
$\tilde{J}(A,B)$ is zero identically. On the other hand,

\begin{eqnarray}&&
\tilde{J}(A,B)
\nonumber\\&&
= \int^{-A}_{- \infty} dw
~\frac{\ln B - \ln \{\alpha w (A + w)\}}{B - \alpha w (A + w)}
+ \int_{-A}^{0} dw
~\frac{\ln B - \ln \{ - \alpha w (A + w)\} + i\pi}{B - \alpha w (A + w)}
\nonumber\\&&
+ \int_{0}^{+ \infty} dw
~\frac{\ln B - \ln \{\alpha w (A + w)\} + 2 i\pi}{B - \alpha w (A + w)}
- i\pi \sum {\rm Res} f(z,A,B).
\nonumber
\end{eqnarray}
\noindent
Thus, changing $w \to - A - w$ in the first integral and $w \to - w$
in the second, we have
\begin{eqnarray}&&\label{int4}
J(A,B)
= \frac{\pi^2}{2\sqrt{\alpha}} B^{-1/2}
\left(1 + \frac{\alpha A^2}{4 B}\right)^{-1/2}
- \frac{1}{2}\int_{0}^{A} dt
~\frac{\ln B - \ln \{\alpha t (A - t)\}}{B + \alpha t (A - t)}
\end{eqnarray}

The roots are contained entirely in the first term on the right of
Eq.~(\ref{int4}), while the logarithms in the second.
The integrals $J_{nm}$ are found by repeated differentiation
of Eq.~(\ref{int4}) with respect to $A,B$. Expanding
the square root $(1 + \alpha A^2/4 B)^{1/2}$ in powers of
$\alpha,$ we find the leading roots needed in Sec.~\ref{loop}
\begin{eqnarray}&&\label{roots}
J^{{\rm root}}_{11} = \frac{\pi^2}{2\sqrt{\alpha}},
~~J^{{\rm root}}_{12} = \frac{\pi^2}{4\sqrt{\alpha}},
~~J^{{\rm root}}_{13} = \frac{3\pi^2}{16\sqrt{\alpha}},
\nonumber\\&&
J^{{\rm root}}_{31} = - \frac{\pi^2}{16}\sqrt{\alpha},
~~J^{{\rm root}}_{32} = - \frac{3\pi^2}{32}\sqrt{\alpha},
~~J^{{\rm root}}_{33} = - \frac{15\pi^2}{128}\sqrt{\alpha}.
\end{eqnarray}
\noindent
Next, expanding the integrand in the second term of Eq.~(\ref{int4}),
we get the leading logarithms
\begin{eqnarray}&&\label{logs}
J^{{\rm log}}_{11} = J^{{\rm log}}_{12} = J^{{\rm log}}_{13} =
- J^{{\rm log}}_{21} =  - J^{{\rm log}}_{22} =  - J^{{\rm log}}_{23} =
\frac{1}{2}\ln\alpha,
\nonumber\\&&
J^{{\rm log}}_{31} = - \frac{\alpha}{4}\ln\alpha,
~~J^{{\rm log}}_{32} = - \frac{\alpha}{2}\ln\alpha,
~~J^{{\rm log}}_{33} = - \frac{3 \alpha}{4}\ln\alpha,
\nonumber\\&&
J^{{\rm log}}_{41} = \frac{\alpha}{12}\ln\alpha,
~~J^{{\rm log}}_{42} = \frac{\alpha}{6}\ln\alpha,
~~J^{{\rm log}}_{43} = \frac{\alpha}{4}\ln\alpha.
\end{eqnarray}

\end{document}